\documentclass[twocolumn,  tighten,  twocolappendix]{aastex63}

\shorttitle{Tidal Features and Properties of Early-type Galaxies}
\shortauthors{Yoon \& Lim}

\begin{document}

\title{Frequency of Tidal Features Correlates with Age and Internal Structure of Early-type Galaxies} 

\email{yyoon@kias.re.kr}

\author[0000-0003-0134-8968]{Yongmin Yoon}
\affiliation{School of Physics, Korea Institute for Advanced Study (KIAS), 85, Hoegiro, Dongdaemun-gu, Seoul, 02455, Republic of Korea}

\author{Gu Lim}
\affiliation{Center for the Exploration of the Origin of the Universe (CEOU),
Astronomy Program, Department of Physics and Astronomy, Seoul National University, 1 Gwanak-ro, Gwanak-gu, Seoul, 08826, Republic of Korea}

\begin{abstract}
Previous studies suggest that compact young early-type galaxies (ETGs) were formed by recent mergers. However, it has not yet been revealed whether tidal features that are direct evidence of recent mergers are detected frequently around compact young ETGs. Here, we investigate how the fraction of ETGs having tidal features ($f_{T}$) depends on age and internal structure (compactness, color gradient, and dust lanes) of ETGs, using 650 ETGs with $M_r\le-19.5$ in $0.015\le z\le0.055$ that are in deep coadded images of the Stripe 82 region of the Sloan Digital Sky Survey. We find that tidal features are more frequent in younger ETGs and more compact ETGs, so that compact young ETGs with ages $\lesssim6$ Gyr have high $f_{T}$ of $\sim0.7$ compared to their less compact or old counterparts with ages $\gtrsim9$ Gyr that have $f_{T}\lesssim0.1$. Among compact young ETGs, those with blue cores have $\sim3$ times higher $f_{T}$ than those with red cores. In addition, ETGs with dust lanes have $\sim4$ times higher $f_{T}$ than those without dust lanes. Our results provide direct evidence that compact young ETGs especially with blue cores and ETGs with dust lanes are involved in recent mergers. Based on our results and several additional assumptions, we roughly estimate the typical visible time of tidal features after a merger, which is $\sim3$ Gyr in the depth of the Stripe 82 coadded images.
\\
\end{abstract}

\section{Introduction}\label{sec:intro}

Early-type galaxies (ETGs) or classical bulges within them are at the final stages of galaxy evolution. The star formation that was once active is no longer active now in ETGs, since they already have consumed available cold gas that is a source of the star formation. Consequently, most of stellar populations in ETGs are old (age $\gtrsim5$ Gyr), so that ETGs generally show red color ($g-r\gtrsim0.7$) in optical bands \citep{Gallazzi2006,Graves2009,Schawinski2014}. ETGs also have centrally concentrated light distributions whose surface brightness profiles can be described by the S\'{e}rsic profiles with high S\'{e}rsic indices of $n\gtrsim3$ or the de Vaucouleurs profile\footnote{Equivalent to the S\'{e}rsic profile with $n=4$.} \citep{Blanton2009,Huertas-Company2013} and usually show round shapes without complex features \citep{Nair2010}. 

One of the widely accepted formation mechanisms of ETGs is a galaxy merger. Mergers between galaxies with enough gas can induce massive star formation \citep{Hernquist1989,Mihos1996,Springel2005,Hopkins2008b,Hopkins2009}, through which available gas is consumed quickly. Feedback effects from strong activities in active galactic nuclei (AGNs) triggered in the merger process can blowout remaining gas or dust \citep{Hopkins2005,Hopkins2008a,Springel2005}. Likewise, mergers can make galaxies quiescent within a short time less than $\sim1$ Gyr in the end \citep{Springel2005,Hopkins2008a,Brennan2015}. At the same time, mergers are also able to cause stellar light distributions in post-merger galaxies to follow the de Vaucouleurs profile \citep{Barnes1988,Naab2006,Hopkins2008b,Hilz2013}. In fact, ETGs, particularly for massive ones, are expected to form and grow through mergers in the $\Lambda$ cold dark matter universe \citep{Baugh1996,Christlein2004,DeLucia2006,DeLucia2007,Wilman2013,Yoon2017,Yoon2019a}, which is a standard model of the universe.

Galaxy mergers are divided into major and minor mergers according to the mass ratio of merging galaxies. Minor mergers are far more frequent at $z<1$ than major mergers \citep{Yoon2017} and main drivers of the morphological evolution of ETGs \citep{Kaviraj2013,Lofthouse2017,Martin2018}. For example, minor mergers with little gas are responsible for size growth of ETGs \citep{Bernardi2011,Oogi2013,Yoon2017}.

Mergers between galaxies produce tidal features such as tidal tails, streams, and shells \citep{Toomre1972,Quinn1984,Barnes1988,Hernquist1992,Feldmann2008}. Thus, tidal features in ETGs are considered the most direct observational evidence for recent mergers. Tidal features around ETGs are frequently detected in deep images. It is known that more than $70\%$ of elliptical galaxies have tidal features in very deep images, although the fraction depends on depth of images \citep{vanDokkum2005,Tal2009,Kaviraj2010,Sheen2012,Kim2013,Hong2015}. 

Using the fact that tidal features are direct traces of recent mergers, many previous studies investigated the correlation between tidal features and properties of ETGs such as AGN activities and color (or age), in order to understand the effect of mergers on the ETGs. For example, \citet{Hong2015} found that almost half of luminous AGN hosts (mostly ETGs) have tidal features. This fraction is significantly higher than that of normal ETGs, suggesting that luminous AGN activities in ETGs are related to galaxy mergers.  

\citet{Schweizer1992} showed that ETGs with younger ages (bluer color at the given luminosity) have more structures produced by mergers. \citet{Tal2009} also found moderate correlation between tidal disturbance (quantitative parameter for tidal features) and broadband color (deviation of $B-V$ at the given luminosity in $V$ band) in a sense that bluer elliptical galaxies have stronger tidal disturbances. Similarly,  \citet{Kaviraj2011} found that morphologically disturbed ETGs have bluer $u-i$ color than relaxed ETGs. In addition, \citet{Schawinski2010} discovered that the fraction of ETGs with tidal features is higher in young ETGs with blue $u-r$ color or residual star formation (traced by emission lines) than in quiescent or old counterparts. These studies imply that young stellar populations in ETGs are connected with recent mergers.

\citet{YP2020} suggested in their study for the fundamental plane (FP) of ETGs that compact young ETGs, especially with blue central color, are likely to have experienced recent (gas-rich) mergers and this is the reason for their large scatter in the FP. This result implies that not only young stellar populations in ETGs, but also compact structures or blue core in ETGs are related to recent mergers, which is also supported by several previous studies \citep{Mihos1994,Rothberg2004,Robertson2006,Hopkins2008b,Hopkins2008c,Hopkins2009}. If this is true, there should be observational evidence that the frequency of tidal features depends on structures of ETGs as well as their ages (or color). However, the correlation between tidal features and the structure of ETGs such as compactness or color gradient has not been intensively studied yet with a large number of ETGs.

Here, we investigate how the fraction of ETGs with tidal features depends on age and internal structure (dust lanes, compactness, and color gradient) of 650 ETGs in the Stripe 82 region of the Sloan Digital Sky Survey (SDSS). By doing so, we discover which ETG populations are directly associated with recent mergers and improve our understanding of the impact of mergers on properties of ETGs. 

 In this paper, we use $H_0=70$ km s$^{-1}$ Mpc$^{-1}$, $\Omega_{\Lambda}=0.7$, and $\Omega_\mathrm{m}=0.3$ as cosmological parameters and the AB magnitude system. 
\\

\section{Sample}\label{sec:sample}
The description for the sample used in this study is similar to that in \citet{YP2020}, except that here we used galaxies in the Stripe 82 region of SDSS. The Stripe 82 region covering $\sim300$ deg$^2$  was scanned $\sim70$--$90$ times in the imaging survey, so that coadded images of Stripe 82 have an $\sim2$ mag deeper detection limit than single-epoch images of SDSS \citep{Jiang2014}. Thus, tidal features of ETGs that are hardly detectable in the SDSS single-epoch images can be discovered in the coadded images of the Stripe 82, as previous studies have shown \citep{Kaviraj2010,Schawinski2010,Hong2015}.

Among the galaxies in the Stripe 82 region, we used galaxies that have spectroscopic redshifts and are classified as ETGs in the KIAS value-added catalog \citep{Choi2010}. This catalog classified galaxies into ETGs or late-type galaxies using photometric parameters such as $u - r$ color, $g-i$ color gradient ($G_{g-i}$), and inverse concentration index ($C_\mathrm{inv}$) in $i$ band \citep[see][]{Park2005}. We note that the completeness and reliability of the classification is $\sim90\%$ \citep{Park2005}. The visual inspection was also performed to increase the accuracy of the classification, and the classifications of $7\%$ of the visually inspected galaxies have been corrected.  For the final ETG sample\footnote{A total of 650 ETGs.} of this study, we examined $fracDev\_r$ from SDSS, which is the weight value of the de Vaucouleurs fit component in the combined model of de Vaucouleurs + exponential disk fit for $r$ band. We found that $70\%$ of ETGs have $fracDev\_r>0.96$ and $94\%$ have $fracDev\_r>0.73$, which means that surface brightness profiles of the ETGs in the final sample are well fitted by the de Vaucouleurs profile and hence they are bulge-dominated galaxies. Therefore, for all magnitudes in this study, we used model magnitudes derived by the de Vaucouleurs profile fits that are from SDSS Data Release 16 \citep[DR16;][]{Ahumada2020}.

We only used ETGs in the low-redshift universe of $z\le0.055$,\footnote{This is the same upper limit as in \citet{YP2020}.} since galaxies at higher redshifts suffer more from the cosmological surface brightness dimming \citep[see Equation 6 in][]{YP2020} and small angular sizes that make the detection of tidal features difficult. The lower limit of the redshift range is set to 0.015 in order to avoid the peculiar velocity effects, which can distort distance-dependent galaxy properties at very low redshifts. The lower limit also prevents the difficulty in classifying tidal features that results from possible problems in the background subtraction caused by large angular sizes of galaxies at very low redshifts. 

The absolute magnitudes in $r$ band were derived by 
\begin{equation}
M_r=m_r -\mathrm{DM}-K_r,
\label{eq:abm}
\end{equation}
where $m_r$ is the apparent magnitude in $r$ band, DM indicates the distance modulus, and $K_r$ is the $k$-correction for $r$ band. The galactic extinctions were corrected in all the apparent magnitudes using the dust maps of \citet{Schlegel1998}. We computed the $k$-correction values using a program of \citet{Blanton2007}, which calculates $k$-correction values by fitting spectral energy distribution (SED) models of diverse metallicities and ages to magnitudes of the five bands ($u$,$g$,$r$,$i$, and $z$) of SDSS. For the SED models, this program uses stellar population models of \citet{Bruzual2003} and the initial mass function of \citet{Chabrier2003}. The $k$-corrections are also applied in $g-r$ color.

We used ETGs with $M_r\le-19.5$ in this study. $M_r=-19.5$ is equivalent to the magnitude limit for spectroscopic samples ($m_r\approx17.77$) at the upper redshift limit of our sample ($z=0.055$). The total number of ETGs with $M_r\le-19.5$ in $0.015\le z\le0.055$ is 1013. We excluded ETGs with small $r$-band axis ratios ($b/a<0.35$) whose photometric properties can be severely affected by internal extinctions. This $b/a$ cut excludes 88 galaxies, so that the number of ETGs with $b/a\ge0.35$ is 925. 

In this study, stellar velocity dispersions of ETGs are used as one of the parameters to compute ages of ETGs. Here, stellar velocity dispersions are aperture corrected to $r_e/8$ by the correction equation in \citet{Jorgensen1995}: 
\begin{equation}
\sigma_0=\sigma_\mathrm{est}\left(\frac{r_\mathrm{fiber}}{r_e/8}\right)^{0.04}, 
\label{eq:vel}
\end{equation}
in which $\sigma_\mathrm{est}$ is the estimated velocity dispersion from SDSS DR16, $r_\mathrm{fiber}$ is the radius of SDSS spectroscopy fibers ($r_\mathrm{fiber}=1.5\arcsec$), and $r_e$ is the angular half-light radius in arcseconds from the de Vaucouleurs fit. It is known that low stellar velocity dispersions less than $\sim100$ km s$^{-1}$ are unreliable \citep{Bernardi2003,Saulder2013} due to  the instrumental limitation of the SDSS spectrograph: the spectroscopic sampling (instrumental dispersion) of the SDSS spectra is 69 km s$^{-1}$ pixel$^{-1}$, and the resolution of the galaxy spectra calculated from the stellar template spectra is $\sim90$ km s$^{-1}$ \citep{Bernardi2003}.  
Thus, we used ETGs with $\sigma_0\ge100$ km s$^{-1}$. Adopting different lower limits of $\sigma_0$ between 70 and 100 km s$^{-1}$ does not alter our main conclusion of this study. Excluding ETGs with $\sigma_0<100$ km s$^{-1}$, the number of ETGs is 701.

Dynamical masses ($M_\mathrm{dyn}$) of ETGs were calculated by 
\begin{equation}
M_\mathrm{dyn}=k\frac{\sigma_0^2R_e}{G}, 
\label{eq:dym}
\end{equation}
where $R_e$ is physical half-light radius of ETGs. In this equation, we use $k=3.8$, which is known to trace the true enclosed mass within $R_e$ \citep{Hopkins2008c}.

As mentioned in Section \ref{sec:tf}, six galaxies among 701 ETGs turn out to be more similar to spiral galaxies. 45 ETGs are located close to bright stars or bright large galaxies that cause high background levels in the coadded images and hence make
it hard to detect tidal features near the ETGs. We excluded these galaxies. Thus, the total number of ETGs in the final sample is 650. 

 \begin{figure}
\includegraphics[width=\linewidth]{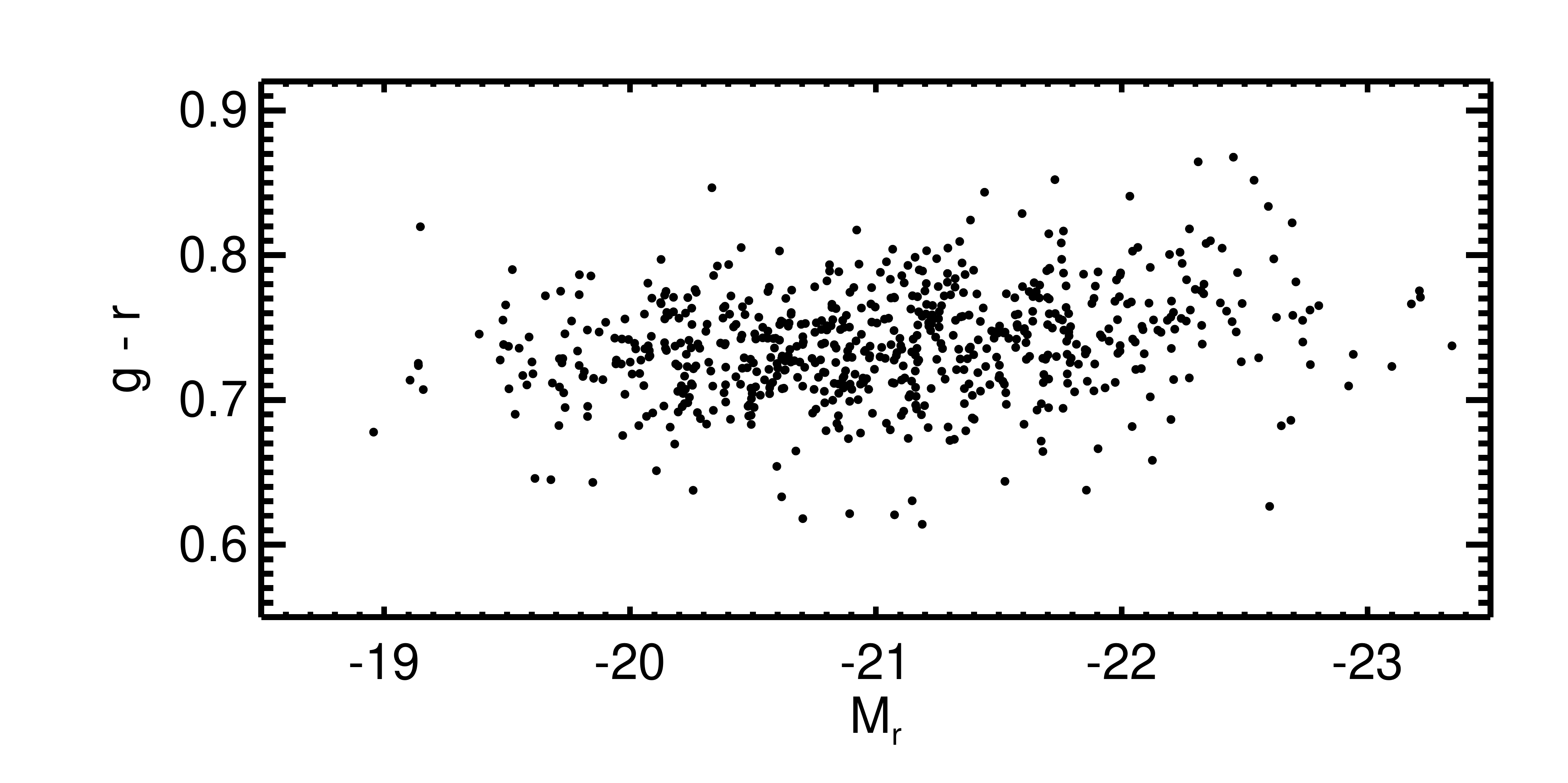}  
\centering
\caption{CMD ($g-r$ vs. $M_r$) for ETGs used in this study. ETGs with $M_r>-19.5$ are not in our final sample, but included in this figure. $g-r$ color values of ETGs are clustered within $\sim0.2$.
\label{fig:cmd}}
\end{figure} 

Figure \ref{fig:cmd} shows the color-magnitude diagram (CMD; $g-r$ vs. $M_r$) for ETGs used in this study. We find that $g-r$ color values of ETGs are clustered within $\sim0.2$, which means that our ETGs are in a tight red sequence. 

 \begin{figure}
\includegraphics[width=\linewidth]{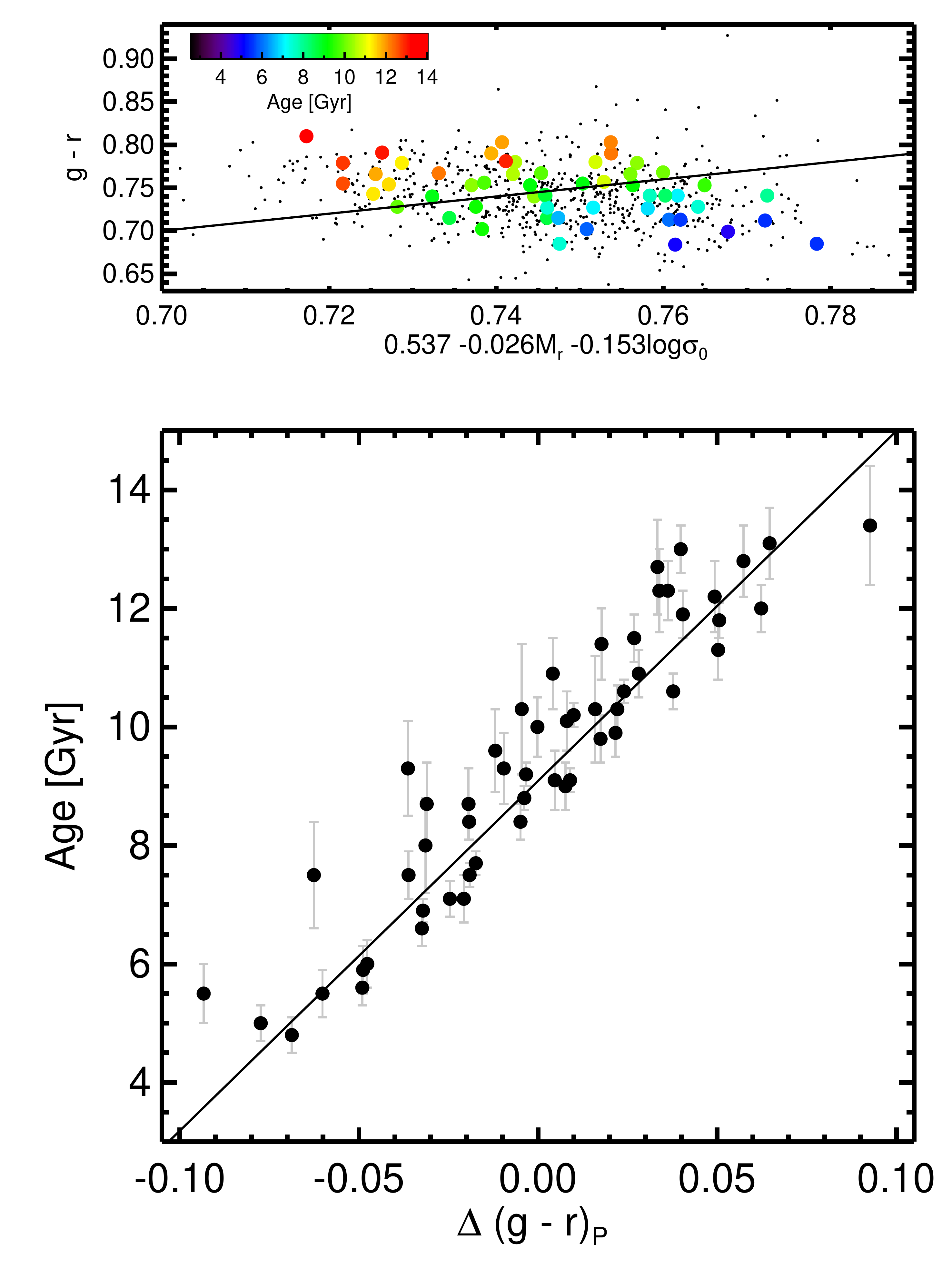}  
\centering
\caption{Top: edge-on view of a plane of constant age (the solid line; Equation \ref{eq:plane}) in three-dimensional parameter space of $g-r$, $M_r$, and $\log\sigma_0$. Also shown are the distributions of our ETGs (black dots) and the groups of quiescent galaxies (colored circles) from \citet{Graves2009}. The stellar population age is color-coded (see the color scale for age). The age varies with the deviation of $g-r$ from the plane ($\Delta (g-r)_\mathrm{P}$). Bottom: age as a function of $\Delta (g-r)_\mathrm{P}$ for the groups of quiescent galaxies (black circles) from \citet{Graves2009}. The gray vertical bars indicate errors in age. The solid line was fitted with the $\chi^2$ minimization method. 
\label{fig:age}}
\end{figure} 

Here, we use a stellar population age indicator for ETGs in \citet{YP2020}. \citet{YP2020} invented an age indicator for ETGs using three parameters ($g-r$, $M_r$, and $\log\sigma_0$), motivated by previous studies showing that a color residual for a given luminosity in CMD correlates with the stellar population age for galaxies in the red sequence \citep{Cool2006,Connor2019} and that both $g-r$ and the stellar velocity dispersion have correlations with age for quiescent galaxies ($g-r$ has a stronger correlation with age than the velocity dispersion; \citealt{Graves2009}). \citet{YP2020} used stacked properties of quiescent galaxies from \citet{Graves2009} to find the age indicator. \citet{Graves2009} divided quiescent galaxies at $0.04<z<0.08$ in SDSS into 54 groups\footnote{Each group contains up to $\sim1000$ galaxies.} according to their $g-r$ color, stellar velocity dispersions, and absolute magnitudes. Then, \citet{Graves2009} derived luminosity-weighted stellar population age from the absorption lines of the stacked spectrum\footnote{The stacked spectra in \citet{Graves2009} have very high signal-to-noise ratios ($\sim100$--$900$).} of each group. 

Using these galaxy groups and their ages, \citet{YP2020} discovered a plane from which deviation in the direction of $g-r$ has the highest correlation with age in the three-dimensional parameter space of $g-r$, $M_r$, and $\log\sigma_0$. This plane was determined by finding a combination of coefficients of the plane equation (e.g., Equation \ref{eq:plane}) that gives the minimum $\chi^2$ in the line fitting which is performed on the relation between ages of the galaxy groups and deviations of $g-r$ from the plane (bottom panel of Figure \ref{fig:age}). The equation of the plane is 
\begin{equation}
(g-r)=0.537-0.026M_r-0.153\log\sigma_0,
\label{eq:plane}
\end{equation}
which corresponds to a constant age of 9.1 Gyr. 

The top panel of Figure \ref{fig:age} shows an edge-on view of the plane (Equation \ref{eq:plane}) in the three-dimensional parameter space. Also shown in the panel are the distributions of our ETGs and the groups of quiescent galaxies from \citet{Graves2009}. This panel indicates that the age of quiescent galaxies changes with the deviation of $g-r$ from the plane of the constant age (hereafter $\Delta(g-r)_\mathrm{P}$). The bottom panel of Figure \ref{fig:age} shows ages as a function of $\Delta(g-r)_\mathrm{P}$ for the groups of quiescent galaxies from \citet{Graves2009}. This panel shows that $\Delta(g-r)_\mathrm{P}$ has a very high correlation with the stellar population age: the linear Pearson correlation coefficient of the two parameters is 0.93. We note that the standard deviation of the age from the fitted line is 0.8 Gyr. 

We use $C_\mathrm{inv}$ from the KIAS value-added catalog as the galaxy structure parameter. $C_\mathrm{inv}$ is inversely proportional to compactness of galaxy light distribution and defined as $R_{p, 50}/R_{p, 90}$, where $R_{p, 50}$ and $R_{p, 90}$ are the seeing-corrected radii at $i$ band containing $50\%$ and $90\%$ of the Petrosian flux, respectively. ETGs generally have $C_\mathrm{inv}\sim0.33$ \citep{Bailin2008}, which is a median value of our ETGs. \citet{Bernardi2003} adopted $C_\mathrm{inv}<0.4$ for one of the criteria to select ETGs. $97\%$ of the ETGs used here have $C_\mathrm{inv}$ values less than 0.4.

$G_{g-i}$ from the KIAS value-added catalog is used for $g-i$ color gradients of galaxies.\footnote{Previous studies \citep{Park2005,Choi2010} used a different expression for color gradients in $g-i$ ($\Delta(g-i)$). However, we use $G_{g-i}$ to avoid confusion with $\Delta(g-r)_\mathrm{P}$, which is the age indicator in this study.}  $G_{g-i}$ is difference in $g-i$ color between the region of $R<0.5R_{p}$ and the annulus region of $0.5R_{p}<R<R_{p}$, in which $R_{p}$ is the Petrosian radius at $i$ band. Negative (positive) $G_{g-i}$ means that the central region has redder (bluer) color than the outer region. 
\\

\section{Detection of Tidal Features}\label{sec:tf}

\begin{figure*}
\includegraphics[width=\linewidth]{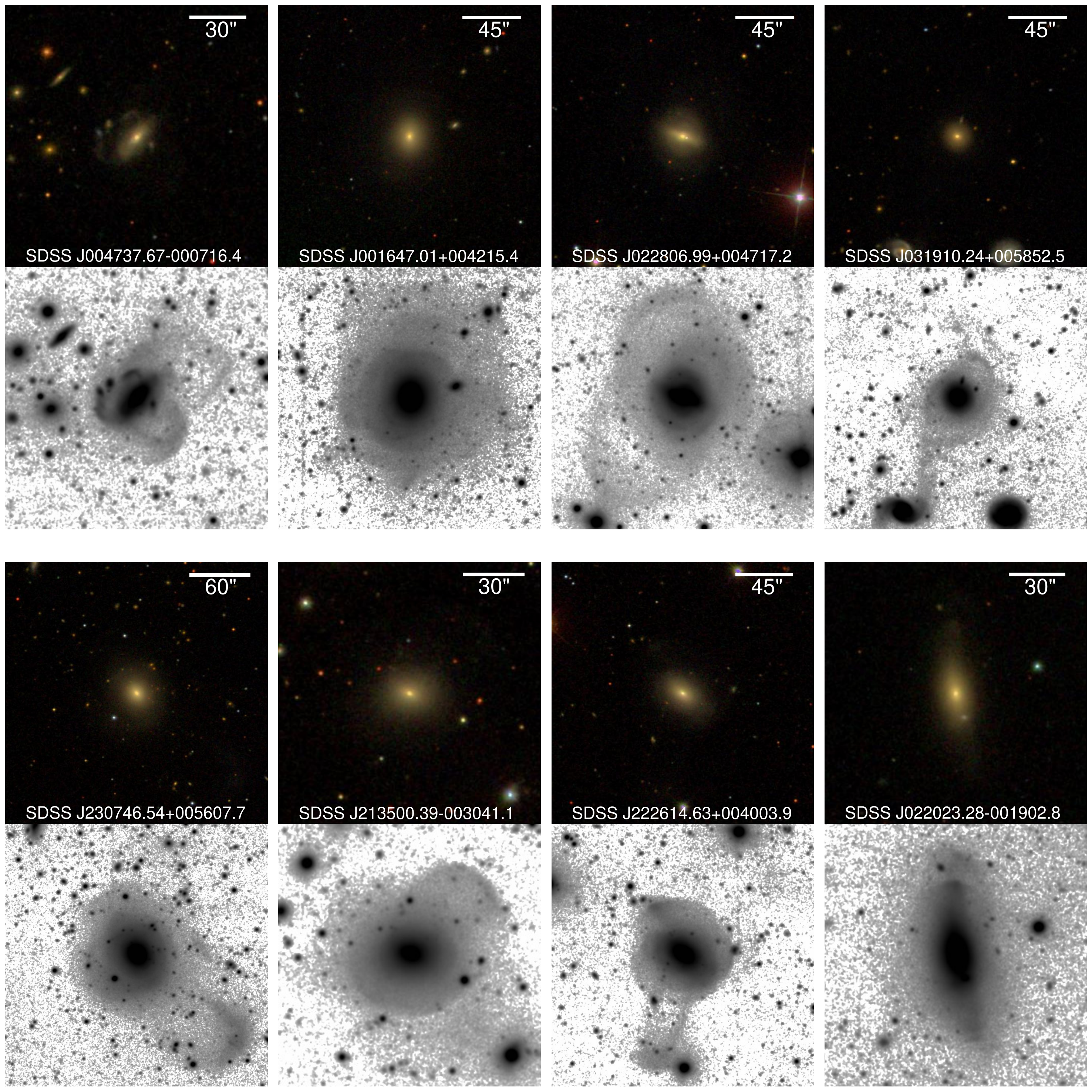}  
\centering
\caption{Color images and deep coadded images of ETGs with tidal features. The horizontal bars in the color images are scales of the images. The name of each ETG is in each color image. We can see very clear tidal features in the deep images that are not visible in the single-epoch color images of SDSS.
\label{fig:ex}}
\end{figure*} 

\begin{figure*}
\includegraphics[width=\linewidth]{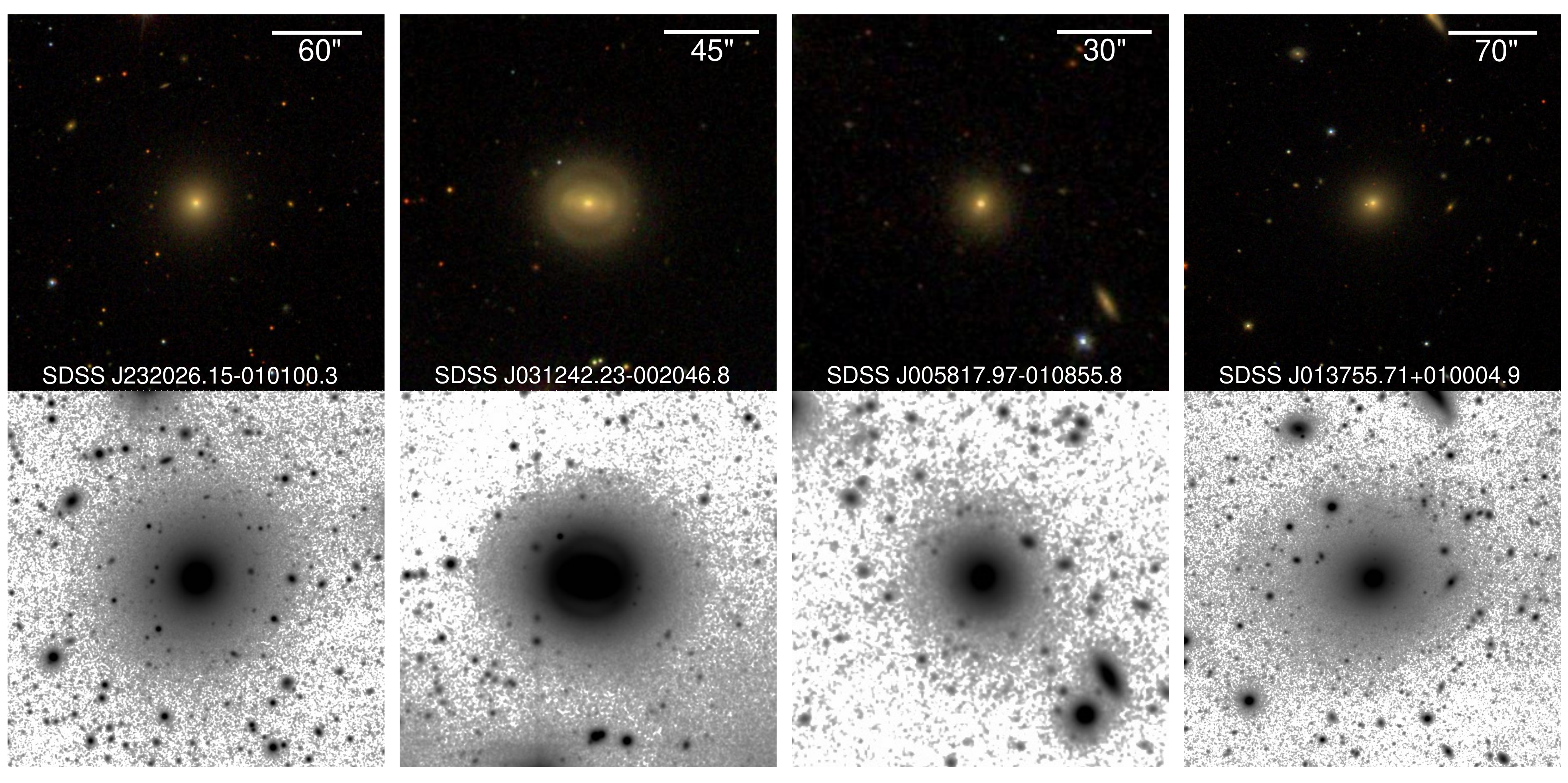}  
\centering
\caption{Color images and deep coadded images of normal ETGs without tidal features. The horizontal bars in the color images are scales of the images. The name of each ETG is in each color image. No tidal feature is visible around ETGs even in the deep images.
\label{fig:ex_normal}}
\end{figure*} 
 
\begin{figure*}
\includegraphics[width=\linewidth]{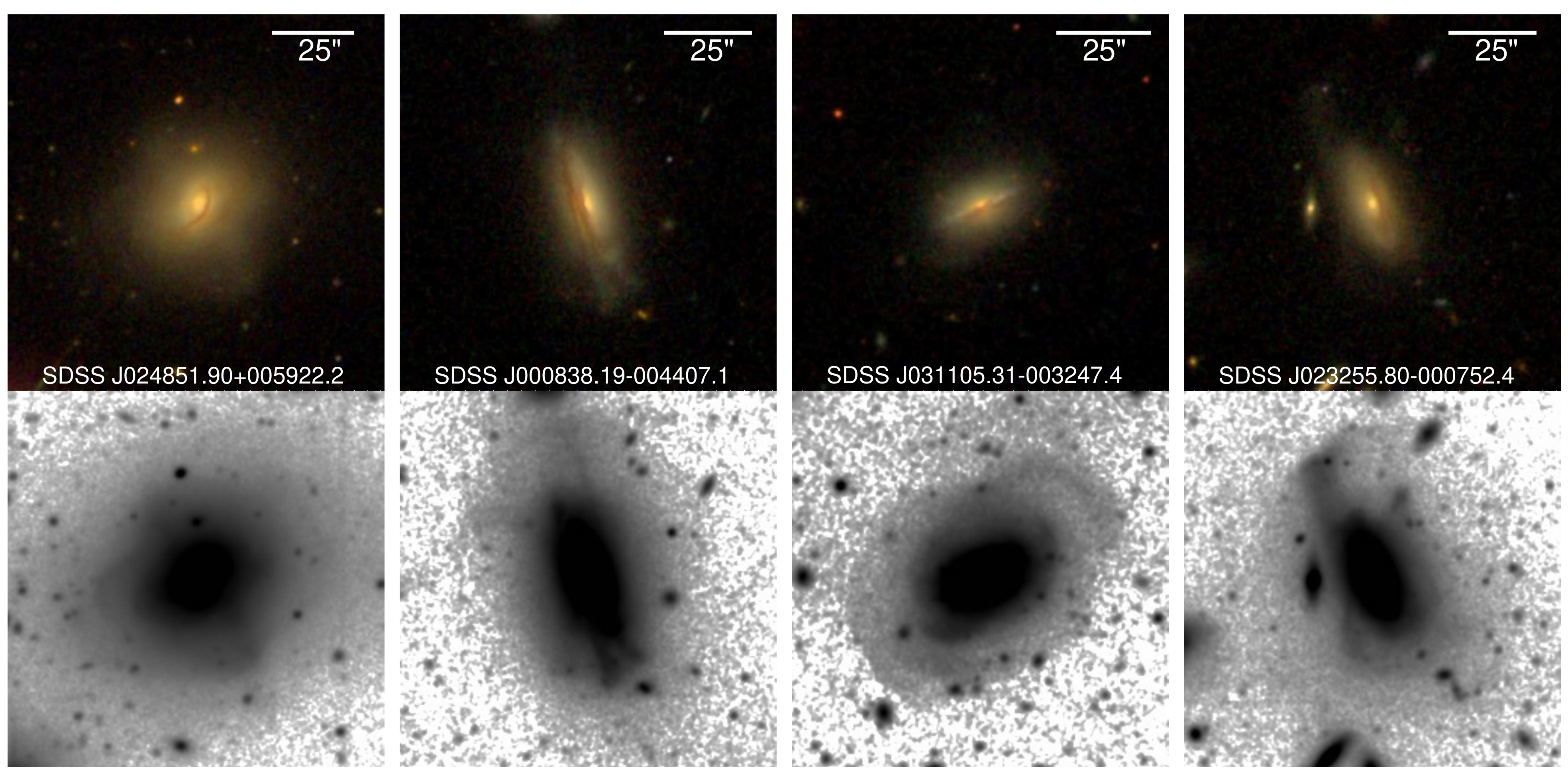}  
\centering
\caption{Color images and deep coadded images of ETGs that have both dust features and tidal features. The horizontal bars in the color images are scales of the images. The name of each ETG is in each color image. We can see obvious dust lanes in the color images and tidal features in the deep images.
\label{fig:ex_dust}}
\end{figure*} 

To detect tidal features of ETGs, we used coadded images of the Stripe 82 from \citet{Jiang2014}, which are $\sim2$ mag deeper than single-epoch images of SDSS. Among the five bands of SDSS, we used $r$-band coadded images whose $5\sigma$ detection limit of the aperture magnitude is 24.6 and surface brightness limit ($1\sigma$ of the background noise over a $1\arcsec\times1\arcsec$ region, hereafter $\mu_\mathrm{limit}$) is $\sim27$ mag arcsec$^{-2}$.  Although some studies used quantitative methods to discover tidal features \citep{vanDokkum2005,Tal2009,Kado-Fong2018}, many studies still prefer visual inspection as it can be difficult to detect faint or irregular tidal features with quantitative methods \citep{Kartaltepe2010,Kaviraj2010,Schawinski2010,Kaviraj2011,Miskolczi2011,Atkinson2013,Hong2015}.  Here, we also performed visual inspection on ETGs to detect tidal features associated with them.

In the process of the visual inspection, we sorted out 45 galaxies that are too close to bright stars or bright large galaxies, since high background levels from the nearby bright sources make it difficult to determine tidal features around these galaxies. We also found six galaxies that are more similar to spiral galaxies than ETGs. We excluded these galaxies in the sample.

  Color images and deep coadded images of several ETGs with tidal features are shown in Figure \ref{fig:ex}. The same images for normal (or relaxed) ETGs that do not have tidal features are shown in Figure \ref{fig:ex_normal}. In Figure \ref{fig:ex}, tidal features that are not apparent in the single-epoch color images are clearly visible around the ETGs in the deep images. On the other hand, normal ETGs without tidal features do not have any unusual features around them even in the deep images in Figure \ref{fig:ex_normal}. We find that 76 ETGs have tidal features among 650 ETGs.

 We also visually sorted ETGs that have clear dust lanes using color images of SDSS. In this study, we treat these ETGs with dust features separately from the other ETGs. Among 650 ETGs, 20 ETGs have obvious dust features. Figure \ref{fig:ex_dust} shows color images and deep coadded images of ETGs that have both dust lanes and tidal features. 

We compared our classifications with those of \citet{Kaviraj2010}. \citet{Kaviraj2010} classified ETGs with $M_r<-20.5$ and $z<0.05$ in the Stripe 82 region into several categories: relaxed ETGs, ETGs with tidal features, and ETGs with dust features. We found that 229 ETGs in our sample are also in the sample of \citet{Kaviraj2010}. In the common ETGs, we detected tidal features in 38 ETGs, among which $89.5\%$ (34/38) are also classified as ETGs with tidal features in \citet{Kaviraj2010}. Two of the four ETGs, which \citet{Kaviraj2010} classified as relaxed ETGs but we defined as ETGs with tidal features, have obvious and prominent tidal features, while the other two ETGs have faint but still discernible tidal features. Reversely, \citet{Kaviraj2010} detected tidal features in 43 ETGs, among which $79.1\%$ (34/43) are also defined as ETGs with tidal features in this study. Three of the nine ETGs, which we classified as normal ETGs but \citet{Kaviraj2010} defined as ETGs with tidal features, do not have any unusual features around them, whereas four of the nine ETGs appear to have possible faint features, but they are so ambiguous that it is difficult to classify them as tidal features with confidence. The other two ETGs are likely to have spurious features generated by bright stars in the vicinity of galaxies. If we correct the definitely misclassified ETGs in \citet{Kaviraj2010}, the percentages mentioned above become $94.7\%$ (36/38) and $85.7\%$ (36/42), respectively, which means that our classifications for tidal features fairly agree with those from \citet{Kaviraj2010}.

 As shown in the comparison with \citet{Kaviraj2010}, we found several ambiguous ETGs for which it is not clear whether or not they have genuine tidal features. These ambiguous cases are not classified as tidal features in this study. However, even if those ambiguous cases are included in the category of tidal features, our results are essentially unchanged.

In the case of dust features in the common ETGs, we found nine ETGs with dust features, among which seven ETGs ($77.8\%$) are also detected as ETGs with dust features in \citet{Kaviraj2010}. We checked the two ETGs in disagreement and found that they have obvious dust lanes. On the other hand, \citet{Kaviraj2010} detected dust features in 10 ETGs, among which seven ETGs ($70.0\%$) are also classified as ETGs with dust lanes in this study. One of the three ETGs in disagreement has no dust feature, while the other two ETGs have possible signs of shaded regions but they are too vague and faint to be guaranteed that they are true dust features. If the definitely misclassified ETGs in \citet{Kaviraj2010} are rectified, the percentages shown above become $100\%$ (9/9) and $81.8\%$ (9/11), respectively, showing that our classifications for dust features are quite consistent with those from \citet{Kaviraj2010}.

In this study, we define the fraction of ETGs with tidal features as
\begin{equation}
f_{T}=N_T/N_\mathrm{ETG},
\label{eq:ft}
\end{equation}
where $N_T$ is the number of ETGs with tidal features and $N_\mathrm{ETG}$ is the number of all types of ETGs (with + without tidal features). As in \citet{Yoon2019b} and \citet{YI2020}, the standard error for the proportion for a binomial distribution is used for the error of $f_{T}$ \citep[see Equation 8 in][]{Yoon2019b}.
\\

\section{Results}\label{sec:results}

\begin{figure*}
\includegraphics[scale=0.285]{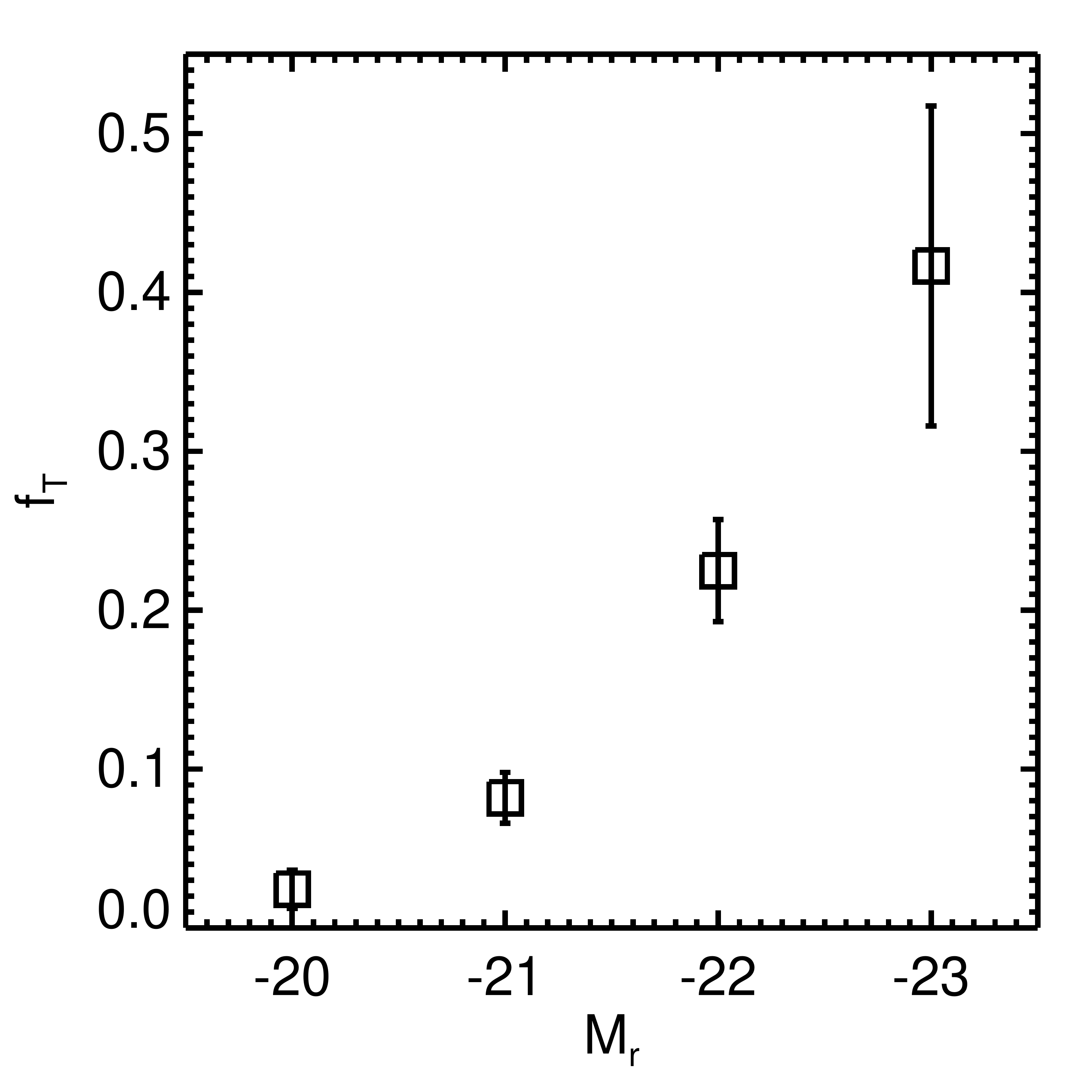} \includegraphics[scale=0.285]{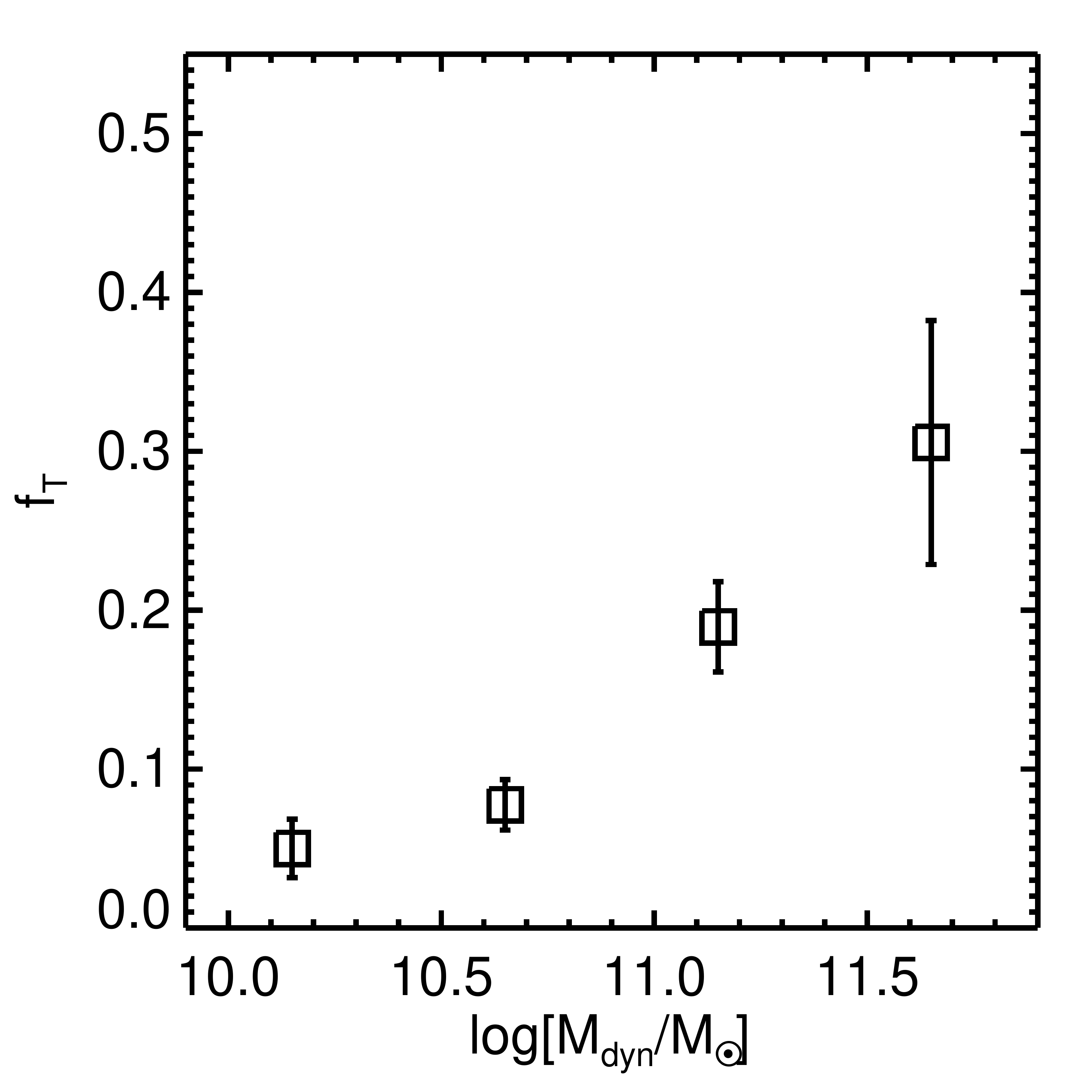}  
\centering
\caption{$f_{T}$ as a function of $M_r$ (left panel) and $M_\mathrm{dyn}$ (right panel). The sizes of bins are 1 mag for $M_r$ and 0.5 dex for $M_\mathrm{dyn}$, respectively. The lowest (highest) bins include all ETGs that have lower (higher) values than the low (high) end of the bins. The figure shows that brighter or more massive ETGs have higher $f_{T}$.
\label{fig:lumass_f}}
\end{figure*} 

\begin{figure}
\includegraphics[width=\linewidth]{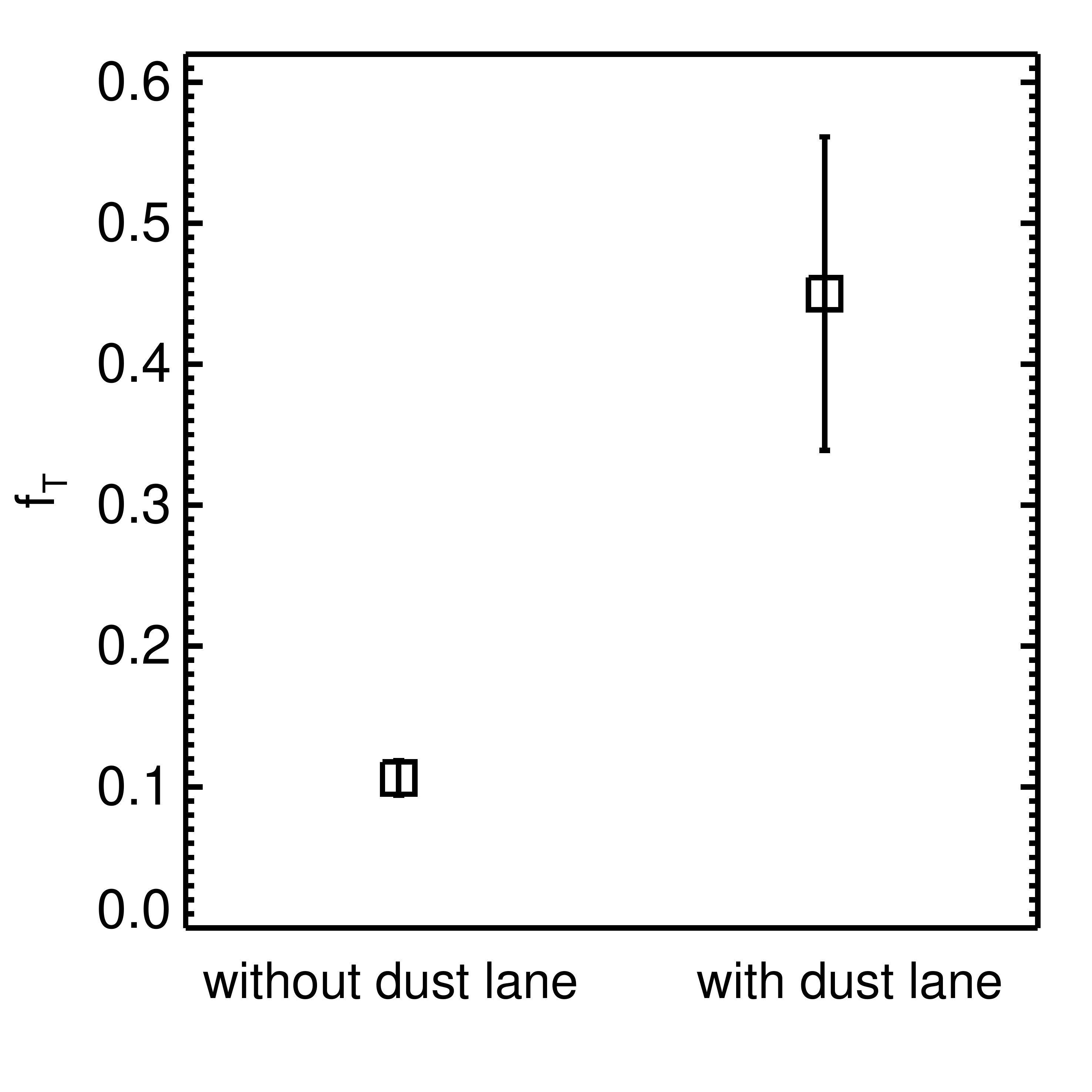} 
\centering
\caption{$f_{T}$ for ETGs with dust lanes (the right square) and without dust lanes (the left square). $f_{T}$ of ETGs with dust lanes is significantly higher than that of ETGs without dust lanes.
\label{fig:dust_f}}
\end{figure} 

Figure \ref{fig:lumass_f} shows how $f_{T}$ changes as a function of $M_r$ (left panel) and $M_\mathrm{dyn}$ (right panel). It is very clear that brighter or more massive ETGs have higher $f_{T}$. Massive ETGs with $\log(M_\mathrm{dyn}/M_\odot)>11.4$ or $M_r<-22.5$ have $f_{T}\sim0.3$--$0.4$, while less massive ETGs with $\log(M_\mathrm{dyn}/M_\odot)<10.4$ or $-20.5<M_r<-19.5$ have $f_{T}\sim0.02$--$0.05$. This result agrees with that of \citet{Hong2015} that $f_{T}$ increases as the bulge luminosity (and the mass of the supermassive black hole) increases. Massive ETGs are on top of the hierarchical assembly of galaxies, which means more massive ETGs are likely to experience more mergers (and hence recent mergers; \citealt{Yoon2017}). Thus, it is natural for more massive ETGs to have higher $f_{T}$. 

In our sample, there are 20 ETGs whose spectra are classified as type 2 AGNs in SDSS based on the optical line ratio diagram of \citet{Baldwin1981}. Their $f_{T}$ is $0.15\pm0.08$ (3/20), which is comparable to $f_{T}$ of the other ETGs (0.12). This implies that type 2 AGNs in ETGs are not related to recent mergers and is consistent with a result of \citet{Schawinski2010}.

We compare $f_{T}$ of ETGs with/without dust lanes in Figure \ref{fig:dust_f}. ETGs with dust lanes have $f_{T}=0.45\pm0.11$. On the other hand, ETGs without dust lanes have $f_{T}=0.11\pm0.01$. Thus, $f_{T}$ of ETGs with dust lanes is $4.2\pm1.2$ times higher than that of ETGs without dust lanes. We note that luminosities of ETGs with dust lanes are on average $0.76$ mag brighter than those without dust lanes. However, this huge difference in $f_{T}$ is not totally caused by the difference in luminosity distributions between the two populations. Even though we make the luminosity distribution of ETGs without dust lanes identical to that with dust lanes by resampling, $f_{T}$ of ETGs without dust lanes is $0.2$, which is still less than half of $f_{T}$ for ETGs with dust lanes. Matching the distributions of other properties such as $\Delta(g-r)_\mathrm{P}$ and $C_\mathrm{inv}$ in the same way as above makes the difference in $f_{T}$ between the two populations even larger. 

Our finding that ETGs with dust lanes have higher $f_{T}$ is consistent with previous studies \citep{Kaviraj2010,Kaviraj2012}, which found that dusty ETGs are often morphologically disturbed. Similarly, morphological disturbances in ETGs with dust lanes are discernible in the single-epoch color images in Figure \ref{fig:ex_dust}. Here, we further confirm that a large fraction of ETGs with dust lanes also have tidal features that can be detected in the deeper images.

\begin{figure*}
\includegraphics[scale=0.285]{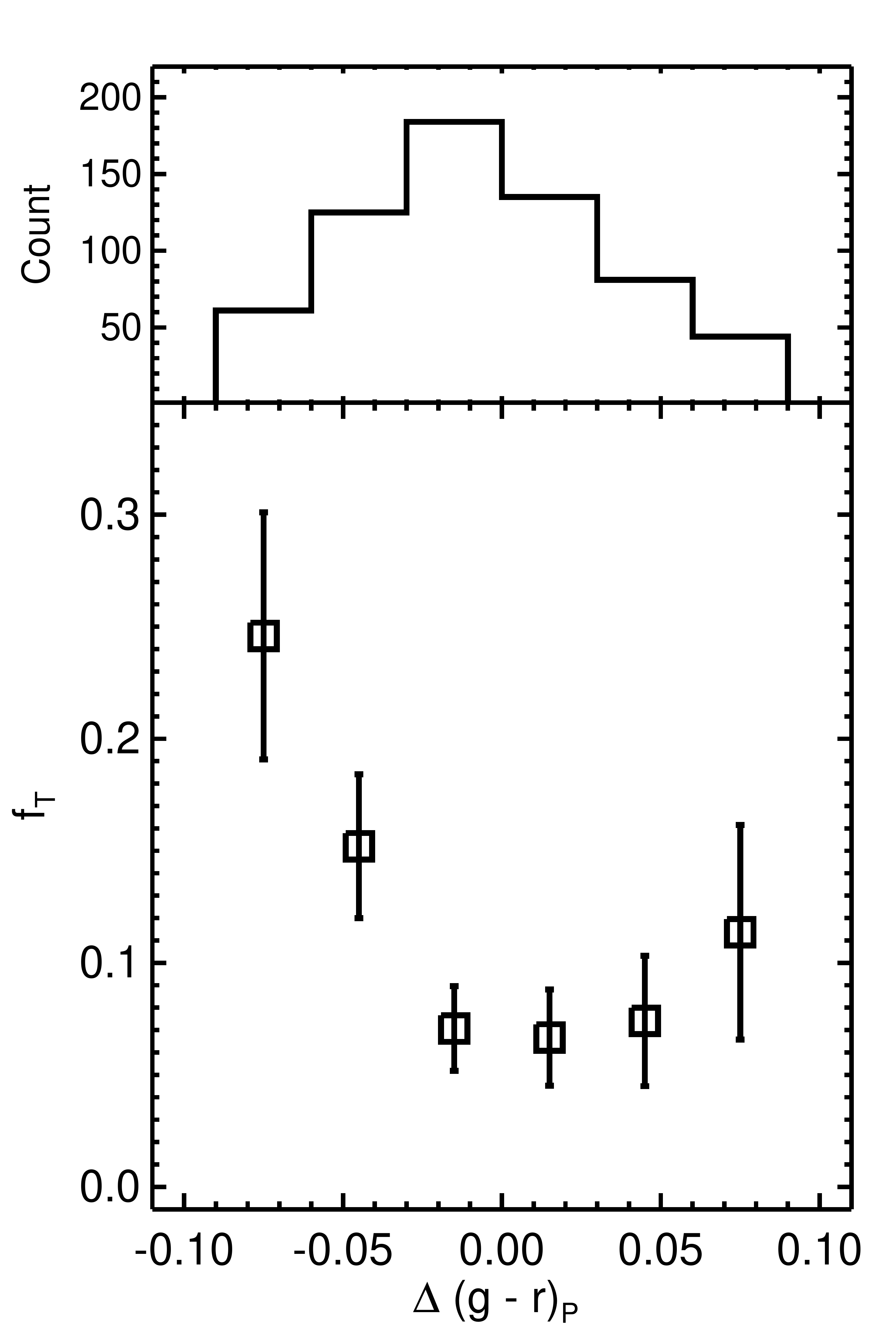} \includegraphics[scale=0.285]{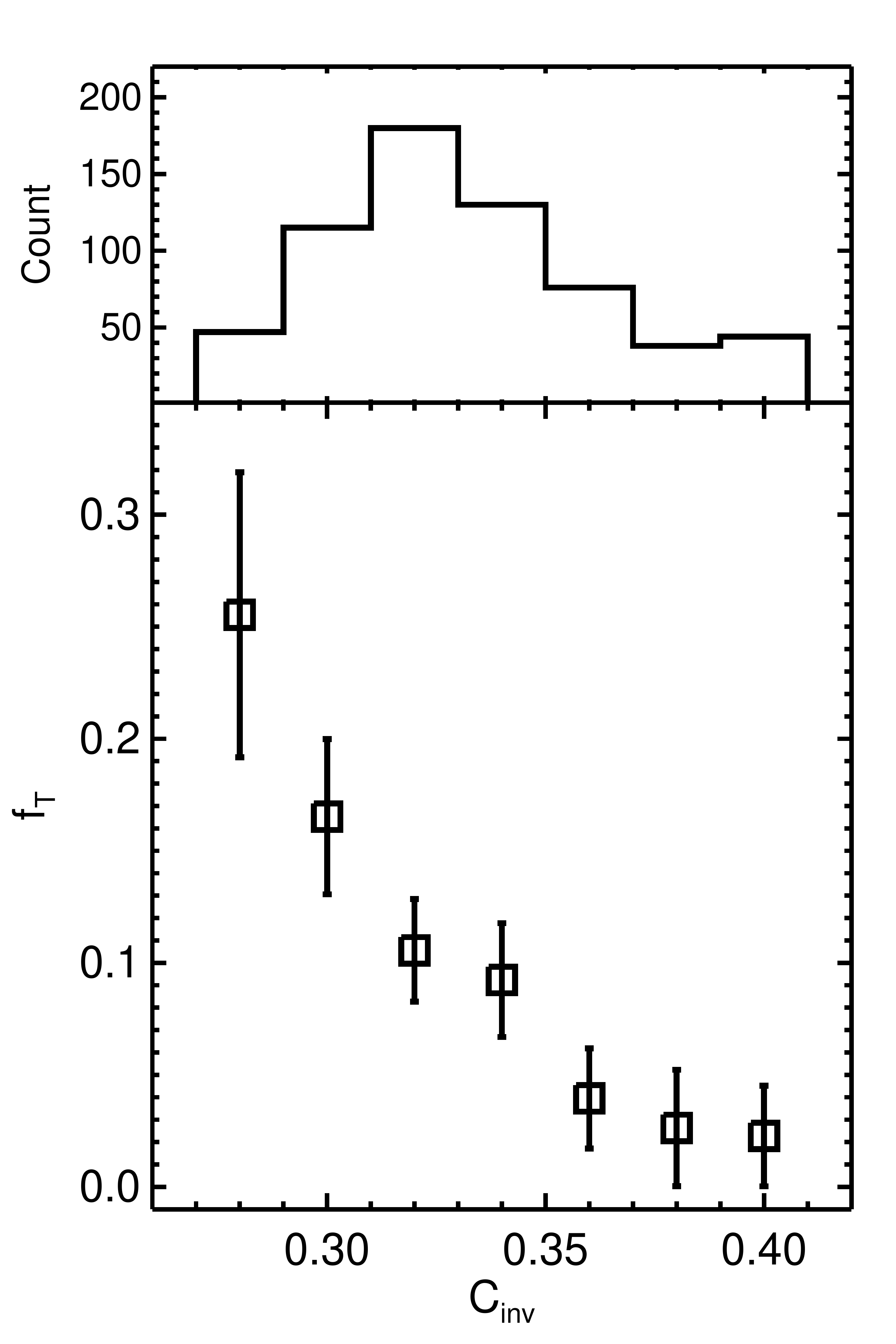}  
\centering
\caption{$f_{T}$ as a function of $\Delta(g-r)_\mathrm{P}$ (left panel) and $C_\mathrm{inv}$ (right panel). The sizes of bins are 0.03 for $\Delta(g-r)_\mathrm{P}$ and 0.02 for $C_\mathrm{inv}$, respectively. The lowest (highest) bins include all ETGs that have lower (higher) values than the low (high) end of the bins. The histograms in the top of each panel show the number of ETGs in each bin. Younger ETGs have higher $f_{T}$ than older ETGs, and more compact ETGs have higher $f_{T}$ than less compact ETGs.
\label{fig:agecon_f}}
\end{figure*} 

\begin{figure*}
\includegraphics[scale=0.285]{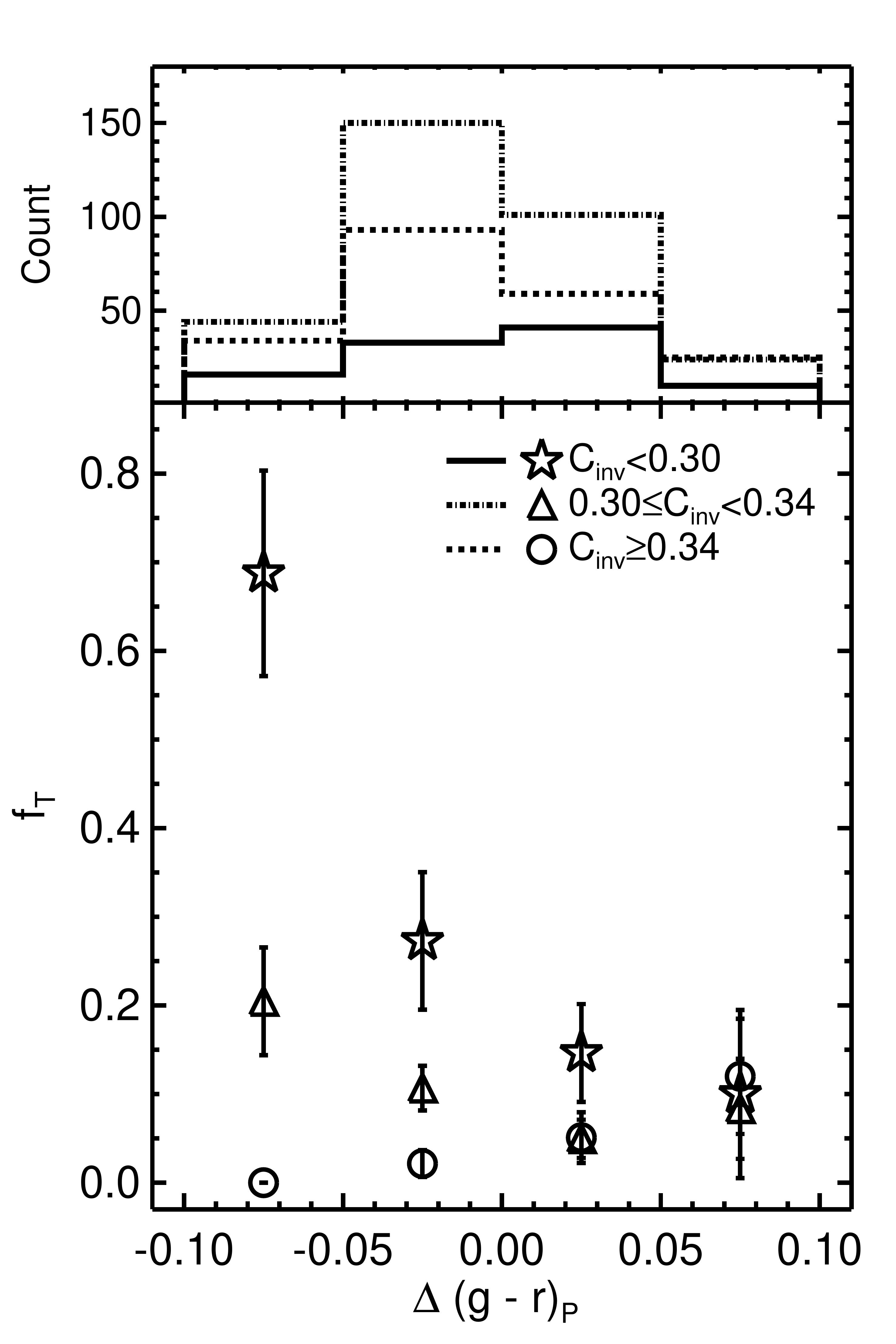} \includegraphics[scale=0.285]{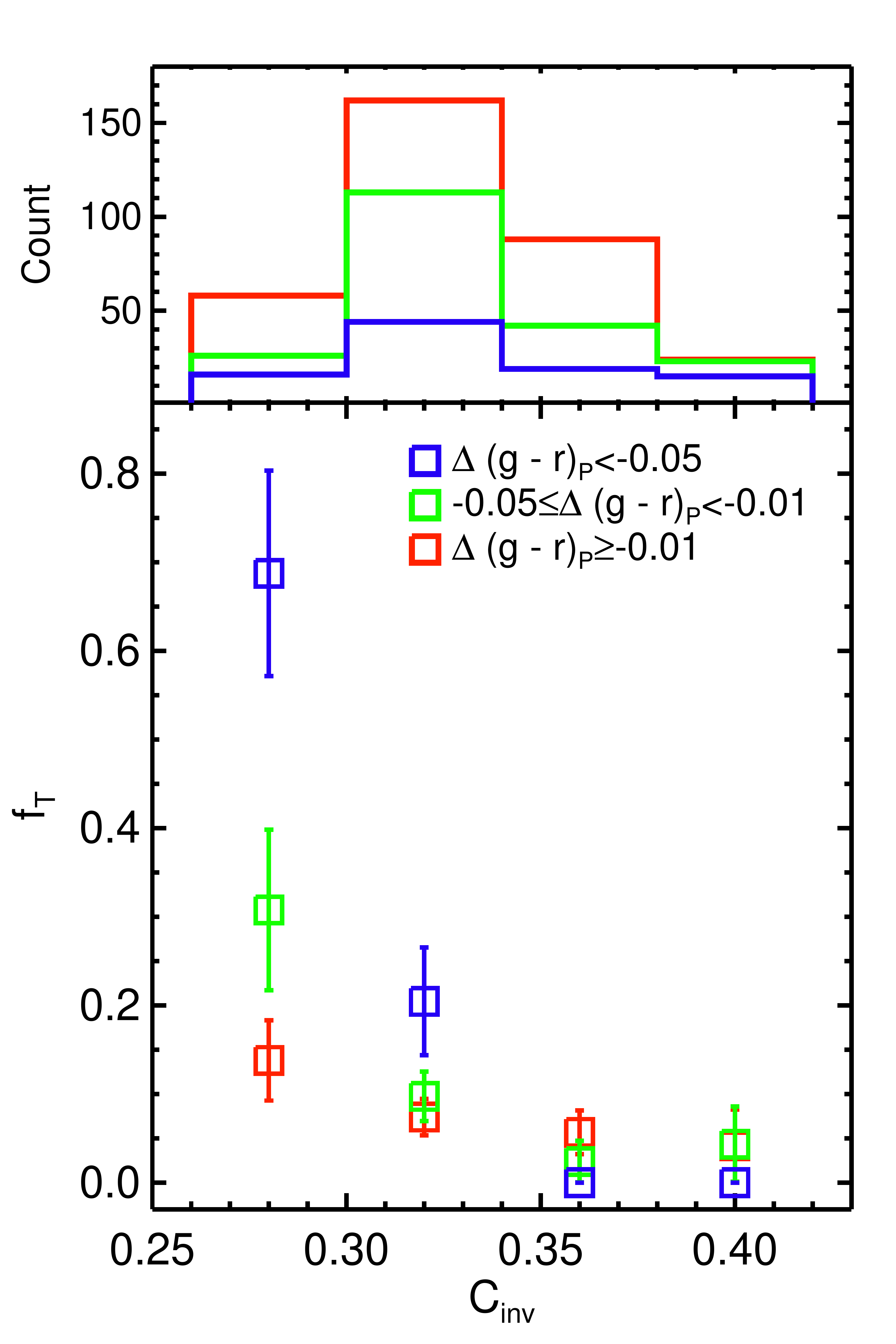}  
\centering
\caption{$f_{T}$ as a function of $\Delta(g-r)_\mathrm{P}$ (left panel) and $C_\mathrm{inv}$ (right panel) for ETGs divided by three bins of $\Delta(g-r)_\mathrm{P}$ or $C_\mathrm{inv}$. The sizes of bins are 0.05 for $\Delta(g-r)_\mathrm{P}$ and 0.04 for $C_\mathrm{inv}$, respectively. The lowest (highest) bins include all ETGs that have lower (higher) values than the low (high) end of the bins. The histograms in the top of each panel show the number of ETGs in each bin. More compact and younger ETGs have higher $f_{T}$ than less compact or older counterparts.
\label{fig:agecon_f2}}
\end{figure*} 

We present results on how $f_{T}$ changes with $\Delta(g-r)_\mathrm{P}$ and $C_\mathrm{inv}$. From now, we exclude 20 ETGs that have dust lanes whose $\Delta(g-r)_\mathrm{P}$ and $C_\mathrm{inv}$ can be largely influenced by their dust lanes. Figure \ref{fig:agecon_f} shows $f_{T}$ as a function of $\Delta(g-r)_\mathrm{P}$ (left panel) and $C_\mathrm{inv}$ (right panel). The left panel of the figure shows that at $\Delta(g-r)_\mathrm{P}\gtrsim-0.03$ (ETGs with age $\gtrsim7$ Gyr), $f_{T}$ is $\sim0.07$ and almost constant within errors. On the other hand, at $\Delta(g-r)_\mathrm{P}\lesssim-0.03$ (ETGs with age $\lesssim7$ Gyr), $f_{T}$ increases with a decrease of $\Delta(g-r)_\mathrm{P}$, so that ETGs with $\Delta(g-r)_\mathrm{P} < -0.06$ (age $\lesssim5.5$ Gyr) have $f_{T}=0.25\pm0.06$. This result shows that younger ETGs have higher $f_{T}$ than older ETGs.

The right panel of Figure \ref{fig:agecon_f} shows that $f_{T}$ increases with a decrease of $C_\mathrm{inv}$, so that ETGs with $C_\mathrm{inv}<0.29$ have $f_{T}=0.26\pm0.06$ while ETGs with $C_\mathrm{inv}\ge0.39$ have $f_{T}=0.02\pm0.02$. This suggests that more compact ETGs have higher $f_{T}$ than less compact ETGs.

In Figure \ref{fig:agecon_f2}, we divide ETGs into three bins of $\Delta(g-r)_\mathrm{P}$ or $C_\mathrm{inv}$ and examine $f_{T}$ as a function of $\Delta(g-r)_\mathrm{P}$ and $C_\mathrm{inv}$. As shown in the left panel of Figure \ref{fig:agecon_f2}, the trend that $f_{T}$ increases with a decrease of $\Delta(g-r)_\mathrm{P}$ is more enhanced for more compact ETGs. Noncompact ETGs with $C_\mathrm{inv}\ge0.34$ do not show such a trend at all (they even have a slightly reversed trend). 

Similarly, another trend that $f_{T}$ increases with a decrease of $C_\mathrm{inv}$ is more intensified for younger ETGs as shown in the right panel of Figure \ref{fig:agecon_f2}. Thus, compact young ETGs with $\Delta(g-r)_\mathrm{P} < -0.05$ and $C_\mathrm{inv}<0.3$ has very high $f_{T}$ ($0.69\pm0.12$). From our results, we conclude that more compact and younger ETGs have higher $f_{T}$ than less compact or older counterparts.

 If we only use massive ETGs with $\log(M_\mathrm{dyn}/M_\odot)\ge10.7$ (median $M_\mathrm{dyn}$ of our sample), $f_{T}$ of each bin in Figures \ref{fig:agecon_f} and \ref{fig:agecon_f2} is increased (errors of $f_{T}$ also rise owing to the small number of ETGs in each bin) and all the trends in the figures are still valid. We note that $86\%$ (12/14) of ETGs with $\log(M_\mathrm{dyn}/M_\odot)\ge10.7$, $\Delta(g-r)_\mathrm{P} < -0.04$, and $C_\mathrm{inv}<0.3$ have tidal features.

\begin{figure}
\includegraphics[width=\linewidth]{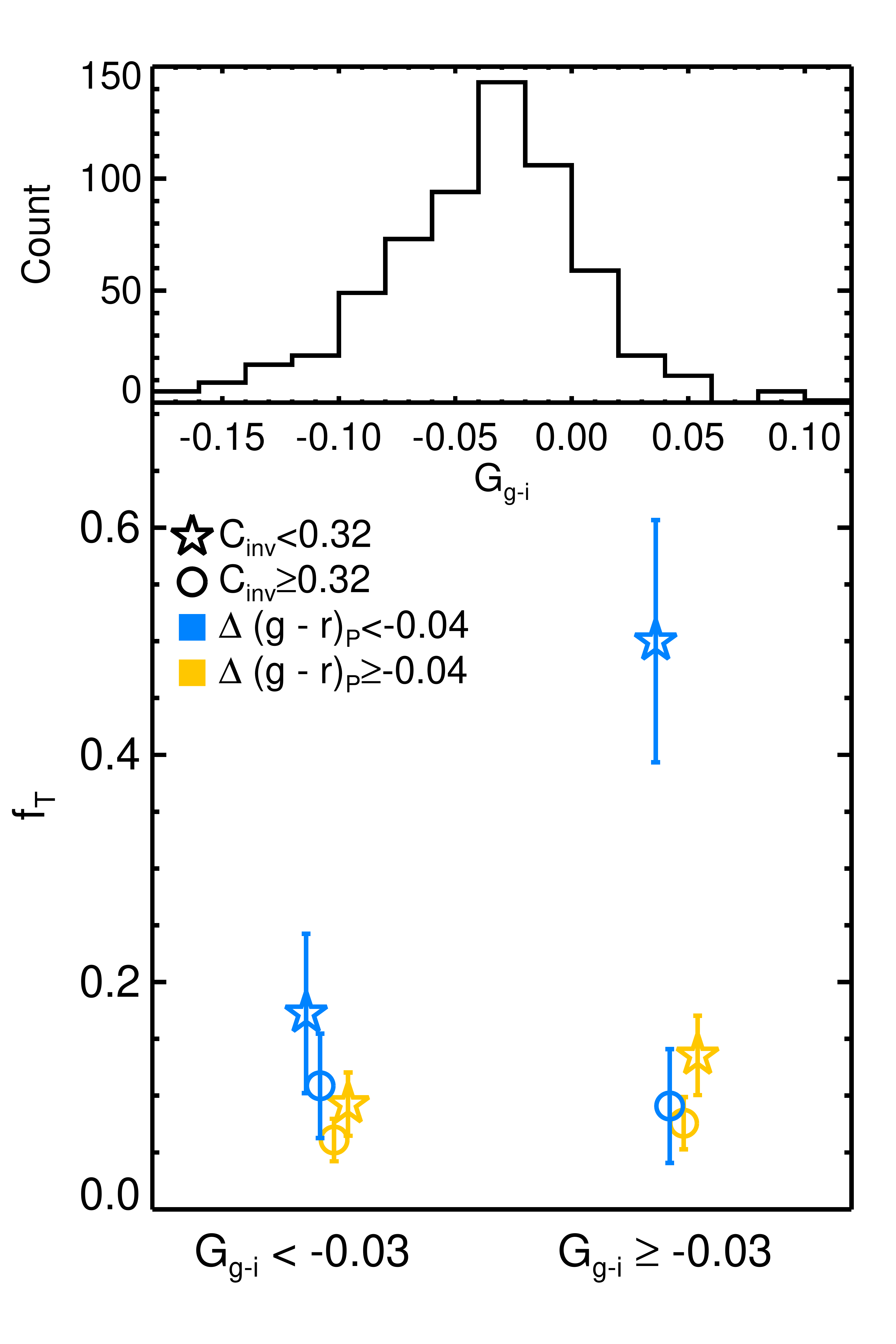} 
\centering
\caption{$f_{T}$ for ETGs with $G_{g-i}<-0.03$ (stars and circles in the left) and $G_{g-i}\ge-0.03$ (stars and circles in the right). Smaller $G_{g-i}$ means redder central color. ETG populations are divided into two bins of $\Delta(g-r)_\mathrm{P}$ or $C_\mathrm{inv}$. The histogram in the top shows the distribution of $G_{g-i}$. Compact young ETGs with bluer central regions have higher $f_{T}$ than the counterparts with redder central regions.
\label{fig:cc_f}}
\end{figure} 

We further examine how $f_{T}$ depends on $G_{g-i}$, which is shown in Figure \ref{fig:cc_f}. In this figure, each ETG population is divided into two $G_{g-i}$ bins: $G_{g-i}<-0.03$ (redder center) and $G_{g-i}\ge-0.03$ (bluer center), in which $G_{g-i}=-0.03$ corresponds to the median $G_{g-i}$ for our ETGs. Figure \ref{fig:cc_f} shows that old ETGs with $\Delta(g-r)_\mathrm{P} \ge -0.04$ have low $f_{T}$ regardless of $G_{g-i}$ ($f_{T}\sim0.11$ for compact ones with $C_\mathrm{inv}<0.32$ and $f_{T}\sim0.07$ for noncompact ones with $C_\mathrm{inv}\ge0.32$). Noncompact young ETGs with $\Delta(g-r)_\mathrm{P} < -0.04$ and $C_\mathrm{inv}\ge0.32$ also have low $f_{T}$ of $\sim0.1$ that does not depend on $G_{g-i}$. However, in the case of compact young ETGs with $\Delta(g-r)_\mathrm{P} < -0.04$ and $C_\mathrm{inv}<0.32$ whose $f_{T}$ is found to be higher than other ETG populations, $f_{T}$ depends on $G_{g-i}$ in such a way that $f_{T}$ at $G_{g-i}\ge-0.03$ ($0.50\pm0.11$) is $2.9\pm1.3$ times higher than $f_{T}$ at $G_{g-i}<-0.03$ ($0.17\pm0.07$). This means that compact young ETGs with bluer central regions have higher $f_{T}$ than the counterparts with redder central regions. We note that $100\%$ (8/8) of ETGs with $\log(M_\mathrm{dyn}/M_\odot)\ge10.7$, $\Delta(g-r)_\mathrm{P} < -0.04$, $C_\mathrm{inv}<0.3$, and $G_{g-i}\ge-0.025$ have tidal features.

Previous studies \citep{Schweizer1992,Tal2009,Schawinski2010} also found evidence for the correlation between color (or age) of ETGs and tidal features (or tidal parameters) as mentioned in Section \ref{sec:intro}. In this study, we newly find that $f_{T}$ correlates not only with age ($\Delta(g-r)_\mathrm{P}$) but also with compactness of structures ($C_\mathrm{inv}$) and color gradients ($G_{g-i}$) of ETGs, and that the correlation between age (or color) and the frequency of tidal features that the previous studies found becomes significantly stronger when only compact ETGs ($C_\mathrm{inv}\lesssim0.3$) are used.
\\

\section{Discussion}\label{sec:discussion}

We detected tidal features in 76 ETGs among 650 ETGs. It is probable that most tidal features detected here were produced by minor mergers. The reason is that ETGs generally experience just one major merger at $z<1$ \citep{Conselice2008,Conselice2009a,Conselice2009b,Yoon2017}, while minor mergers occur far more frequently in ETGs \citep{Yoon2017}.

Our results suggest direct evidence that compact young ETGs with blue cores or ETGs with dust lanes are involved in recent mergers. According to previous studies, the recent mergers that occurred in compact young ETGs are likely to be gas-rich. It is known that a gas-rich merger induces gas inflows into the center of the merger remnant owing to loss of angular momentum via tidal torques and radiation during the merger process, and then a starburst is triggered in the central region \citep{Hernquist1989,Barnes1991,Barnes1996,Hopkins2008c}. This process produces compact central light components of blue young stellar populations with typical sizes of $\sim0.5$--$1$ kpc in the inner regions of post-merger galaxies \citep{Mihos1994,Robertson2006,Hopkins2008b,Hopkins2008c,Hopkins2009,Kormendy2009}.

This is also supported by a recent work in \citet{YP2020}, who used previous simulation studies about remnants of gas-rich mergers \citep{Hopkins2008b,Hopkins2009} and generated a simple model of a typical galaxy that experienced a gas-rich merger, in order to examine properties of the post-merger galaxy as a function of time \citep[Figure 12 of][]{YP2020}. They showed that compact young ETGs with blue cores can be produced by the recent gas-rich mergers and their compact structures do not change significantly with the passage of time, so that they naturally become compact old ETGs in the end. This scenario can give the reason why $f_{T}$ decreases in compact old ETGs (the left panel of Figure \ref{fig:agecon_f2}), since tidal features around ETGs settle down and are not visible in the deep images after enough time. 

It is known that dust lanes in ETGs can be generated in recent gas-rich merger processes \citep{Oosterloo2002,Clemens2010,Kaviraj2012,Shabala2012}. Thus, recent mergers that occurred in ETGs with dust lanes are also likely to be gas-rich.

We find that $\sim7$--$10\%$ of old ETGs with age $\gtrsim7$ Gyr ($\Delta(g-r)_\mathrm{P}\gtrsim-0.03$) also have tidal features. Since the old ETGs are numerous (see the histogram the left panel of Figure \ref{fig:agecon_f}), this means that a considerable number of tidal features are also detected in old ETGs, although $f_{T}$ is low. Taking into account that dry mergers (mergers without abundant gas) are not able to noticeably make age of ETGs young, tidal features in ETGs with old stellar populations seem to be produced by dry mergers. 

The left panel of Figure \ref{fig:agecon_f2} shows that $f_{T}$ of compact ETGs with $C_\mathrm{inv}<0.30$ decreases and converges to a low value of $\sim0.1$ as $\Delta(g-r)_\mathrm{P}$ increases. Although many different merger scenarios are possible, if we assume a single and typical gas-rich merger model (scenario) in \citet{YP2020} for the formation of all the compact ETGs, it is possible to roughly estimate the typical detectable or visible time of tidal features after a merger (hereafter, $t_d$) in the depth of the Stripe 82 images. Here, we define $t_d$ as the time taken for $f_{T}$ to fall to half its initial value. We assume that almost all very early tidal features after the merger are strong enough to be detected in the depth of the Stripe 82 images. Then, $t_d$ is the time when $f_{T}$ becomes $\sim0.55$, considering that $10\%$ of ETGs are assumed to have tidal features from dry mergers as mentioned above. Performing simple interpolation in Figure \ref{fig:agecon_f2}, $f_{T}$ for compact ETGs is $\sim0.55$ at $\Delta(g-r)_\mathrm{P}\sim-0.06$, which corresponds to an age of $\sim5.5$ Gyr. According to the typical gas-rich merger scenario in \citet{YP2020}, the light-weighted age of $\sim5.5$ Gyr for post-merger galaxies corresponds to $\sim3$ Gyr since the merger\footnote{Specifically, since the starburst triggered by the merger.} happened \citep[see Figure 12 of][]{YP2020}, which implies that $t_d$ is $\sim3$ Gyr in the images with $\mu_\mathrm{limit}\sim27$ mag arcsec$^{-2}$. We note that our estimation for $t_d$ is only meaningful in that it gives an approximate value for $t_d$ based on observational data, since compact ETGs in reality can be formed by various merger scenarios.

Despite such a crude estimation for $t_d$, our estimation of $\sim3$ Gyr is roughly consistent with previous numerical simulations on the visible time of tidal features. For example, \citet{Ji2014} found that $t_d$ is $\sim1$--$2$ Gyr in shallow depth images with $\mu_\mathrm{limit}=25$ mag arcsec$^{-2}$, while in deeper images with $\mu_\mathrm{limit}=28$ mag arcsec$^{-2}$, $t_d$ can be increased up to $\sim5$ Gyr. \citet{Mancillas2019} also suggested $t_d\sim2$--$4$ Gyr in deep images with $\mu_\mathrm{limit}\geq29$ mag arcsec$^{-2}$.
\\

\section{Summary}\label{sec:summary}
We investigate how the fraction of ETGs having tidal features ($f_{T}$) correlates with age and internal structure (dust lanes, compactness, and color gradient) of ETGs, using 650 ETGs with $M_r\le-19.5$ in $0.015\le z\le0.055$ in the Stripe 82 region of SDSS. We define $\Delta(g-r)_\mathrm{P}$ in the three-dimensional parameter space of $g-r$, $M_r$, and $\log\sigma_0$ and use it as an age indicator of ETGs. We use $C_\mathrm{inv}$ for structure parameter of ETGs that is inversely proportional to compactness of galaxy light distribution. For color gradient of ETGs, we use difference in $g-i$ color between the inside and the outside of the galaxy ($G_{g-i}$). We performed visual inspection to detect tidal features using deep coadded images of the Stripe 82 region that are $\sim2$ mag deeper than usual single-epoch images of SDSS. Since ETGs generally experience far more minor mergers than major mergers at $z<1$, it is probable that most of tidal features detected in this study were produced by minor mergers. Our main conclusions are as follows.

\begin{enumerate}
\item Tidal features are more frequent in more massive or brighter ETGs: massive ETGs with $\log(M_\mathrm{dyn}/M_\odot)\ge11.4$ have $f_{T}=0.31$, while less massive ETGs with $\log(M_\mathrm{dyn}/M_\odot)<10.4$ have $f_{T}=0.05$.

\item ETGs with dust lanes have four times higher $f_{T}$ than those without dust lanes: ETGs with dust lanes have $f_{T}=0.45$, while ETGs without dust lanes have $f_{T}=0.11$.

\item Tidal features are more frequent in younger ETGs: young ETGs with age $\lesssim5$ Gyr have $f_{T}\sim0.25$, while old ETGs with age of $\gtrsim7$ Gyr have $f_{T}\sim0.07$. Moreover, tidal features are also more frequent in more compact ETGs: compact ETGs with $C_\mathrm{inv}<0.29$ have $f_{T}=0.26$, while noncompact ETGs with $C_\mathrm{inv}\ge0.39$ have $f_{T}=0.02$.

\item Compact young ETGs ($C_\mathrm{inv}<0.30$ and age $\lesssim6$ Gyr) have very high $f_{T}$ of $\sim0.7$, compared with their less compact or old counterparts with age $\gtrsim9$ Gyr that have $f_{T}\lesssim0.1$. Furthermore, among compact young ETGs, those with blue central regions ($G_{g-i}\ge-0.03$) have three times higher $f_{T}$ than those with red central regions ($G_{g-i}<-0.03$).

\item Our results give direct evidence that compact young ETGs (particularly with blue cores) and ETGs with dust lanes are associated with recent (gas-rich) mergers.

\item Using our results and several assumptions, we roughly estimate that the typical visible time of tidal features after a merger is $\sim3$ Gyr in the depth of the Stripe 82 coadded images. This estimation is consistent with previous studies based on numerical simulations.
\end{enumerate}

Fainter and more tidal features should be detected in much deeper images than the Stripe 82 coadded images used here. So, it would be interesting to see whether our results shown here are still valid through the deeper images in the future. 

Tidal features can be divided into several types such as tidal tails, streams, and shells. According to previous studies \citep{Feldmann2008,Tal2009,Mancillas2019}, different types of tidal features may differ in origin and lifetime, which implies that such different types of tidal features may have different correlations with properties of their host ETGs. Thus, we expect to conduct a similar study on diverse types of tidal features using a larger ETG sample in the future.

\acknowledgments
This work was supported by a KIAS Individual Grant PG076301 at the Korea Institute for Advanced Study. GL acknowledges to support from the National Research Foundation of Korea (NRF) grant, No. 2020R1A2C3011091, funded by the Korea government (MSIT).

\end{document}